
\input mtexsis
\paper
\twelvepoint
\Eurostyletrue
\superrefsfalse
\def\x{{ \bf X }}
\def\yo1{{f_\pi ^2}}
\def\nc{{{\cal N}_c}}

\def\coset{{ {\cal G}\over{\cal H} }}

\def\pmu{{ \partial_{\mu} }}
\def\pmuu{{ \partial^{\mu} }}

\def\dilf{{ \chi }}

\def\glke{{ {\rm G}_{\mu\nu}^{a} {\rm G}^{\mu\nu}_{a} }}
\def\roke{{ {\rm G}_{\mu\nu}{\rm G}^{\mu\nu} }}
\def\lam1{{\int_0^\infty}}

\def\ga{{ g_{A} }}

\def\ssd{{ \sigma ^2+ \vec\pi ^2 }}

\referencelist
\reference{mgeo} A.~Manohar and H.~Georgi, \journal Nucl. Phys. B;234,189
                (1984)
\endreference
\reference{sw4} S.~Weinberg, \journal Phys. Rev. Lett.;65,1181 (1990)
\endreference
\reference{bls} W.~Broniowski, M.~Lutz, and A.~Steiner,
                \journal Phys. Rev. Lett.;71,1787 (1993)
\endreference
\reference{pdr}  S.~Peris and E.~de Rafael,
                 \journal Phys. Lett. B;309,389 (1993)
\endreference
\reference{sp1} S.~Peris, \journal Phys. Lett. B;268,415 (1991)
\endreference
\reference{sw5} S.~Weinberg, \journal Phys. Rev. Lett.;67,3473 (1991)
\endreference
\reference{*sw5a} D.A.~Dicus, D.~Minic, U.~van Kolck, and R.~Vega,
                 \journal Phys. Lett. B;284,469 (1992)
\endreference
\reference{sw6} S.~Weinberg, \journal Physica;96A,327 (1979)
\endreference
\reference{ew}    E.~Witten, \journal Nucl. Phys. B;160,57
                (1979)
\endreference
\reference{sw2} S.~Weinberg, \journal Phys. Rev. ;177,2604 (1969)
\endreference
\reference{*sw2a} S.~Weinberg, \journal Phys. Rev. Lett.;65,1177 (1990)
\endreference
\reference{dm} J.L.~Gervais and B.~Sakita,
                  \journal Phys. Rev. Lett.;52,87 (1984)
\endreference
\reference{*dma} R.~Dashen and A.~Manohar,
                   \journal Phys. Lett. B;315,425 (1993)
\endreference
\reference{shur} E.V.~Shuryak, \journal Nucl. Phys. B;203,93,116 (1982)
\endreference
\reference{je} J.~Ellis, \journal Nucl. Phys. B;22,478 (1970)
\endreference
\reference{bz}  B.~Zumino, Lectures on Elementary Particles and Quantum
                         Field Theory, edited by S. Deser et al.
                    (M.I.T. Press 1970)
\endreference
\reference{cw} S.~Coleman and E.~Weinberg,
               \journal Phys. Rev. D;7,1888 (1973)
\endreference
\reference{geo} H.~Georgi, \journal Nucl. Phys. B;331,311 (1990)
\endreference
\reference{us} S.~Weinberg, ``Unbreaking Symmetries,'' to be published in
                 ``Festschrift for Abdus Salam''
\endreference
\reference{hk} T.~Hatsuda and T.~Kunihiro,
\journal Phys. Rev. Lett.;55,158 (1985)
\endreference
\reference{gock} A.~Gocksch, \journal Phys. Rev. Lett.;67,1704 (1991)
\endreference
\reference{bb} M.C.~Birse and M.K.~Banerjee,
               \journal Phys. Lett. B;136,284 (1984)
\endreference
\reference{*bba} P.~Jain, R.~Johnson, and J.~Schechter
               \journal Phys. Rev. D;38,1571 (1988)
\endreference
\reference{fa} See, for example, S.~Kuyucak and A.~Faessler,
               \journal Phys. Lett. B;169,128 (1986)
\endreference
\reference{ccj} C.G. Callan, S. Coleman, and R. Jackiw,
                        \journal Ann. of Phys;59,42 (1970)
\endreference
\endreferencelist
\titlepage
\obeylines
\hskip4.3in{CPP-94-2}
\hskip4.3in{DOE/ER/40427-01-N94}
\unobeylines
\title
The Dilated Chiral Quark Model
\endtitle
\author
S.~R.~Beane\footnote\dag{{\rm sbeane@utpapa.ph.utexas.edu}}
Center for Particle Physics
Department of Physics
The University of Texas at Austin
Austin, TX 78712
\endauthor
\author
U.~van Kolck\footnote\S{{\rm vankolck@alpher.npl.washington.edu}}
Department of Physics, FM-15
University of Washington
Seattle, WA 98195
\endauthor
\abstract
It is argued that constituent quarks live in an effective theory that
possesses an approximate conformal invariance. An effective lagrangian
is constructed which in the large-$\nc$ approximation incorporates
Regge asymptotic constraints. The resulting picture explains why
linear-sigma models provide successful constituent quark descriptions,
both at zero and finite temperature.  Our analysis suggests an
interesting relation between non-linearly realized conformal
invariance and the completion of chiral multiplets in the broken
symmetry phase.
\vskip2.2in
\center{To be published in Physics Letters B}\endcenter
\endabstract
\vfill\eject                                    
\singlespaced
Particle physicists have long puzzled over why the non-relativistic
quark model works so well for hadrons made up of light quarks.
Although the emergence of QCD has not resolved this conundrum, a
partial explanation can be found in the chiral quark model of Manohar
and Georgi\ref{mgeo}. If one assumes that the chiral symmetry breaking
energy scale is larger than the confinement scale, then the effective
low-energy field theory between these scales consists of the
interactions among constituent quarks, gluons, and {\it elementary}
Goldstone bosons of chiral symmetry.  Because they are now massive and
interact via weak gluon exchange (which accounts for spin-dependent
corrections), the constituent quarks can be treated approximately as
non-relativistic particles.  It remains unexplained, however, why
constituent quarks behave as bare Dirac particles, with axial vector
coupling $\ga$ close to (but smaller than) one and no anomalous
magnetic moment, $\mu _q$. Some time ago Weinberg\ref{sw4} suggested
that this could be understood by imposing on the chiral quark model
the requirement of Regge asymptotic behavior.  He argued that in the
large-$\nc$ limit $\ga=1$ and $\mu _q$=0$^1$\vfootnote1{ This
conclusion has been, however, challenged recently in \Ref{bls} (See
also \Ref{pdr}).}.  Various authors then showed that leading $1/\nc$
corrections push $\ga$ down but leave $\mu _q$
unchanged\ref{sp1}\ref{sw5}.  In this letter we concentrate on the
asymptotic behaviour of the $\pi$-$\pi$ interaction in the chiral
quark model.  (The importance of this process in establishing the
leading $1/\nc$ corrections to $\ga$ has been emphasized in
\Ref{pdr}.)  We pursue the following approach: we take the chiral
quark model seriously as an effective theory of QCD, and consider the
constraints that follow from imposing Regge asymptotic behaviour.

Our starting point is an old ``theorem''of Weinberg\ref{sw6} which
states that the content of quantum field theory is given entirely by
general physical principles like unitarity and cluster decomposition,
together with the assumed internal symmetries. QCD is characterized by
$SU(3)_{c}$ gauge invariance and (in the chiral limit to which we will
restrict ourselves in this letter) a global $SU(2)_{L}\otimes
SU(2)_{R}$ symmetry, which is however not obviously manifest in the
spectrum.  Its usual representation employs current quark fields that
realize chiral symmetry linearly, but have non-trivial bilinear vacuum
expectation values. Weinberg's ``theorem'' suggests that an
``equivalent'' field theory exists, which employs constituent quark
fields that realize chiral symmetry non-linearly and therefore
includes explicitly also Goldstone boson fields.  Since these
additional degrees of freedom destroy the good asymptotic behaviour
that one would expect of QCD, this effective description is useful
only at low-energies; the introduction of additional degrees of
freedom is necessary to ensure true QCD behavior. Furthermore, it is
not unreasonable to suppose that these additional degrees of freedom
have, as do pions, an origin in the symmetries of QCD. We will show
that conformal invariance emerges as an important symmetry in this
respect.

Since the chiral quark model is a non-renormalizable field theory,
quanta not present in the low-energy theory should appear
at higher energies in order to effect the necessary cancellations
among graphs.
This expectation can be made practical by combining low-energy
theorems of chiral symmetry with
dispersion sum rules (particularly in the $\pi$-$\pi$ sector where the
amplitude at threshold is completely determined by the geometry of the
coset space) that find their
justification in Regge asymptotic behaviour.
In the large-$\nc$
approximation, one would expect that sum rules for Goldstone boson
scattering off a quark or meson target are saturated by single
particle states in the narrow width approximation\ref{ew}.
The sum rules can then be shown to possess algebraic
content\ref{sw2}.  We will briefly review this formalism.

Chiral symmetry severely constrains low-energy scattering processes
with Goldstone bosons on external lines, and allows an expansion of
amplitudes in powers of momenta. What is less well-known is that one
can also obtain interesting constraints by expanding chiral tree
amplitudes in inverse powers of momenta. In this case, the parameters
associated with the heavy particle content contribute to the
power of energy which is protected by chiral symmetry, and for which
there is therefore no invariant counterterm. The algebraic
consequences of chiral symmetry arise from the need for cancellations
among these heavy-particle contributions. (For higher powers of energy
there exist chiral invariant contact interactions with arbitrary
coefficients, and therefore no constraints are obtained.)  The
complete set of Adler-Weisberger sum rules can be put into the form

$$\lbrack \x_{i},\x_{j}\rbrack= i\epsilon _{ijk}T_{k}, \EQN aw $$ where
$T_{i}$ denotes the isospin generator,

$$\lbrack T_{i},T_{j}\rbrack= i\epsilon _{ijk}T_{k}, \EQN id $$
and $\x_{i}$ is the isovector, reduced matrix element for emission of a pion
with isospin $i$ between one particle states. Hence,

$$\lbrack T_{i},\x_{j}\rbrack=
i\epsilon _{ijk}\x_{k}. \EQN xd $$
\Eqs{aw} through \Ep{xd} imply
that particles appearing in chiral tree-graphs furnish representations
of $SU(2)\otimes SU(2)$. In general, strongly interacting particles
would be expected to fall into complicated reducible representations.
One can learn more by considering sum rules
of superconvergence type.
The absence of Regge trajectories with ``exotic'' quantum numbers
leads to sum rules which can be expressed in algebraic form as

$$\lbrack \x_{j},\lbrack M^{2},\x_{i}\rbrack\rbrack =-\lbrack
M_{4}^{2}\rbrack\delta _{ij}. \EQN sc $$
These sum rules reveal that
the diagonal-mass-squared matrix can be written as
${M}^{2}$=${M}_{0}^{2}$+${M}_{4}^{2}$, where ${M}_{0}^{2}$ is a chiral
singlet, and ${M}_{4}^{2}$, the symmetry breaking part of the mass
matrix, transforms as the fourth component of a chiral four-vector.
An important consequence of this sum rule is that
the masses of any two particles in an irreducible representation are
equal.

In \Ref{sw4} Weinberg considered the representation involving the
constituent quark; \Eqs{aw} through \Ep{sc} then constrain $\ga$.
Sum rules for pions scattering off baryon targets are readily
constructed, yielding commutation relations for baryon states similar
to the Gervais-Sakita-Dashen-Manohar large-$\nc$ consistency
conditions\ref{dm}: for a baryon $T_{i}^{N}=\sum_{a=1}^N T_{i(a)}=
O(1)$, and $\x_{i}^{N}=\sum_{a=1}^N \x_{i(a)}= N (\x_{i}^{N(0)}+ 1/N
\x_{i}^{N(1)}+ O(1/N^{2}))$ with $\x_{i}^{N(0,1)}=O(1)$, so \Eq{aw}
gives $\lbrack \x_{i}^{N(0)},\x_{j}^{N(0)}\rbrack=0$, $\lbrack
\x_{i}^{N(0)},\x_{j}^{N(1)}\rbrack +\lbrack
\x_{i}^{N(1)},\x_{j}^{N(0)}\rbrack =0$, and so on. Consistency
conditions for the baryon mass matrix follow in a similar fashion from
\Eq{sc}. (The reason behind the breaking of the chiral symmetric structure for
baryons in the $\nc\rightarrow\infty$ limit is unclear to us.)

We now turn to the $\pi$-$\pi$ sector in the chiral quark model.
Since confinement is a large-separation effect, we will assume that
$q\bar q$ states have no place in the effective theory$^2$\vfootnote2{
This approach is consistent with the philosophy of the instanton model
of the QCD vacuum, which provides interesting support for the
two-scale hypothesis. See, for example, \Ref{shur}.}.
This assumption eliminates several chiral quark model paradoxes. If
$q\bar q$ states are a part of the spectrum, then there is a
pseudoscalar $q\bar q$ bound state as well as an elementary
pion\ref{mgeo}.  Furthermore, there is a duality between skyrmion
stability and the excitation(s) that unitarize the $\pi$-$\pi$
interaction. That is, vector mesons like $\rho$, which couples
strongly to the pion, are responsible for the stability of skyrmions.
In the chiral quark model, the existence of stable skyrmions would
lead to a second double-counting problem.  Therefore, consistency
itself suggests that mesonic states are generated by the running of
$\alpha _s$ at the confinement scale. We will see that this simple
point has far reaching consequences.

If there are no color-singlet excitations in the effective theory with
non-trivial transformation property under the unbroken isospin
subgroup, then $\pi$ must be joined by {\it at least one}
isoscalar in a representation of the form $\bf 4\oplus\bf
1\oplus...\oplus\bf 1$.  We can go further by considering the
representation involving the pion in the color-singlet sector.  There a
quartet consisting of $\pi$, $\epsilon$, and the longitudinal
components of $\rho$ and $a_1$ can be shown to fill out the reducible
representation given by $\bf 4\oplus\bf 6$ of chiral $SU(2)\otimes
SU(2)$\ref{sw2}.  Consistency suggests that above the confinement
scale, the pion falls into the simplest irreducible representation,
the $\bf 4$, in which a single scalar
acts as the fourth component of a chiral four vector. According
to \Eq{sc}, there can be no mass splitting within the multiplet, and
hence the scalar should be more or less degenerate with the pion.

Since all quanta in the effective theory are accounted for, the
interactions among the constituent quarks, pions, and the scalar
should be weak enough that no bound states appear in the spectrum.
Hence one might expect the existence of an additional symmetry that
keeps the interactions involving the scalar under control.
One might also suspect that such a symmetry is
spontaneously broken, since Goldstone bosons are the only ``natural''
massless scalars.  Fortunately, such a symmetry exists. The QCD
lagrangian, in the absence of large quantum effects (for example, at a
fixed point of the $\beta$-function), possesses an approximate
conformal invariance, which if relevant below the chiral symmetry
breaking scale is necessarily spontaneously broken.
The resulting elementary dilaton field is the
natural chiral partner of the elementary pion.
This approach has the virtue of allowing a
systematic effective lagrangian analysis, including symmetry breaking,
based on the operator scaling properties of QCD.  Furthermore, there
is the appealing feature that all the quanta in the effective theory,
besides the QCD degrees of freedom, find their origin in symmetry.

With this in mind, consider the coset space,

$${\coset}= {{SU(2)_{L}\otimes SU(2)_{R}\otimes{\cal
C}}\over {SU(2)\otimes{\cal P}}}. \EQN cqm1 $$
$\cal C$ denotes the conformal group, and $\cal P$ the Poincar\'e group.
(We also assume invariance under the discrete symmetries of QCD.)
The Goldstone triplet $\pi$ is contained in a matrix field $U=\xi\xi$,
$\xi= exp ({{i\pi_{i}T_{i}}\over{f_{\pi}}})$
which transforms
under $SU(2)_{L}\otimes SU(2)_{R}$ as
$\xi\rightarrow L\xi v^{\dagger}=v\xi R^{\dagger}$,
where $v$, a non-linear function of $L$, $R$, and $\pi$, is implicitly
defined.
The condition
that a chiral transformation be conformally invariant requires that
the Goldstone boson fields have scale dimension,
$d_{U}$=0.
Out of $\xi$ one can construct
$V_{\mu}={1\over 2}(\xi ^{\dagger}\pmu\xi +\xi\pmu\xi^{\dagger});
  V_{\mu}\rightarrow vV_{\mu}v^{\dagger}+v{\partial _{\mu}}v^{\dagger}$, and
$A_{\mu}={1\over 2}i(\xi ^{\dagger}\pmu\xi -\xi\pmu\xi^{\dagger});
  A_{\mu}\rightarrow vA_{\mu}v^{\dagger}$.
We introduce a constituent quark doublet,
which transforms as
$\psi\rightarrow v\psi$
under $SU(2)_{L}\otimes SU(2)_{R}$, and as
$\psi\rightarrow e^{d_{\psi}\lambda}\psi$
under dilatations. The scale dimension $d_{\psi}$ has no invariant
significance when dilatation symmetry is spontaneously broken\ref{je}, and
so choice of $d_{\psi}$ corresponds to the definition of the constituent
quark field. It is convenient to
introduce the dilaton field $\sigma$, which scales
inhomogeneously, by way of a
scalar field $\tilde\dilf$=$\exp ({{\sigma}\over {f_d}})$
with canonical scale dimension, $d_{\tilde\dilf}$=1.

The most general dilatation invariant chiral quark
lagrangian$^3$\vfootnote3{ Meaning of course a lagrangian that yields
a dilatation invariant action.}
at leading order is then given by
\offparens
$$
\eqalign{
{\cal L}=&{\bar\psi}i(\slashchar D+\slashchar V)\psi\tilde\dilf ^{-2d_{\psi}+3}
            +\ga \bar\psi\slashchar A\gamma _{5}\psi\tilde\dilf ^{-2d_{\psi}+3}
            -{m}\bar\psi\psi\tilde\dilf ^{-2d_{\psi}+4}
        +{1\over 4}\yo1 tr({\pmu}U{\pmuu}U^{\dagger}){\tilde\dilf ^2}\cr
            &+{1\over 2}f_d^2 \pmu\tilde\dilf\pmuu\tilde\dilf
            -{1\over{2}}tr(\roke )
+ig_{s}(\bar\psi\gamma_{\mu}\psi )(\pmuu\tilde\dilf)\tilde\dilf ^{-2d_{\psi}+2}
 -V(\tilde\dilf)+...,\cr }
\EQN dqm1 $$\autoparens
where
$m$ is the constituent quark mass
and the covariant derivative is given by $D_{\mu}=\pmu +igG_{\mu}$, with
$G_{\mu}$ the gluon field of field strength $G_{\mu\nu}$.
Although the lagrangian, \Eq{dqm1}, is manifestly scale invariant, the
derivative interactions are not
necessarily conformally invariant.
It is not difficult to show that conformal invariance determines
$g_{s}=-{(d_{\psi}-{3\over 2})}$\ref{je}. Clearly the
canonical choice $d_{\psi}={3\over 2}$ is most convenient. We then have,
defining $\dilf\equiv{f_d\tilde\dilf}$ and $\kappa\equiv{\yo1\over{f_d^2}}$,
\offparens
$$
\eqalign{
{\cal L}=& {\bar\psi} i(\slashchar D+\slashchar V)\psi +\ga
\bar\psi\slashchar A\gamma _{5}\psi
-\sqrt{\kappa}{m\over{f_{\pi}}}\bar\psi\psi\dilf +{1\over 4}\kappa
tr({\pmu}U{\pmuu}U^{\dagger}){\dilf ^2}\cr &+{1\over
2}\pmu\dilf\pmuu\dilf -{1\over{2}}tr(\roke ) -V(\dilf)+... \cr} \EQN
dqm2 $$\autoparens
In the absence of explicit dilatation breaking, the
potential $V(\dilf)$ is a flat function, and therefore the dynamical
mechanism responsible for fixing the vacuum expectation value of
$\dilf$ is absent$^4$\vfootnote4{ Note that although $\dilf ^4$ looks
conformally invariant, stability of the vacuum requires subtracting
$4{f_d^3}\sigma$, which is clearly not invariant.}. Thus, conformal
symmetry is spontaneously broken only if it is explicitly
broken\ref{bz}. Hence we must include minimal symmetry breaking.  In
QCD, breaking of conformal invariance is governed by the trace of the
energy-momentum tensor: the divergence of the dilatation current
$J_{\mu}$ is

$$\pmuu J_{\mu}=\Theta _{\mu}^{\mu}={{\beta (g)}\over{2g}}\glke \EQN qcd $$
in the chiral limit.
If there are low-energy effective theories of QCD that possess an
approximate conformal invariance, then the symmetry breaking terms in the
effective theory should transform like
$\lambda \pmuu J_{\mu}$, where $\pmuu J_{\mu}$ is the most general
dimension-four, gauge invariant Lorentz scalar constructed out of the
effective fields. Stability of the vacuum implies

$$V(\dilf)=-{ {\kappa m_{\sigma}^2}\over{8\yo1} }
\biggl\lbrack{1\over 2}\dilf ^4
-\dilf ^4\log({ \kappa{\dilf ^2}\over{{\yo1}}})\biggr\rbrack,
\EQN pot $$
where $m_{\sigma}$ is the dilaton mass.
There is of course an infinite tower of additional symmetry breaking
terms involving the interactions of the dilaton with itself and with
the other degrees of freedom in the effective theory, but these are
expected to be suppressed by powers of the dilaton mass divided by the
chiral symmetry breaking scale.

How do we put the pions and the dilaton in a ``Goldstone quartet''?
We can transform our effective theory to a linear basis, $\Sigma =
U\dilf\sqrt{\kappa}$ where $\Sigma\equiv{\sigma '}
+i{\vec\tau}\cdot{\vec\pi '}$,
and
$Q={1\over 2}((\xi +\xi ^{\dagger})+\gamma _{5}(\xi - \xi
^{\dagger}))\psi$.  In terms of the new fields the lagrangian, \Eq{dqm2},
 is given by

\offparens
$$
\eqalign{
&{\cal L}=i{\bar Q}\slashchar D Q
     -{1\over{2}}tr(\roke )
           +{1\over 4}tr(\pmu\Sigma\pmuu\Sigma ^{\dagger})
       -{m\over{2f_{\pi}}}{\bar Q}\lbrack\Sigma +\Sigma ^{\dagger}
   -\gamma _{5}(\Sigma -\Sigma ^{\dagger})\rbrack Q\cr
 &+{{(1-\kappa )}\over{16\kappa}}tr(\Sigma\Sigma ^{\dagger})^{-1}
 tr(\pmu(\Sigma\Sigma ^{\dagger}))tr(\pmuu(\Sigma\Sigma ^{\dagger}))\cr
 &-{i\over 4}(1-\ga)tr(\Sigma\Sigma ^{\dagger})^{-2}{\bar Q}\lbrack
 tr(\slashchar\partial(\Sigma\Sigma ^{\dagger}))
   \lbrace\Sigma ,\Sigma ^{\dagger}\rbrace -2tr(\Sigma\Sigma ^{\dagger})
   (\Sigma\slashchar\partial\Sigma ^{\dagger}+
    \Sigma ^{\dagger}\slashchar\partial\Sigma)\rbrack Q \cr
 &-{i\over 2}(1-\ga)tr(\Sigma\Sigma ^{\dagger})^{-1}{\bar Q}\lbrack
   {\gamma _{5}}(\Sigma\slashchar\partial\Sigma ^{\dagger}-
    \Sigma ^{\dagger}\slashchar\partial\Sigma)\rbrack Q \cr
&+{{m_{\sigma}^2}\over{64\kappa\yo1}}
\lbrack tr(\Sigma\Sigma ^{\dagger}) \rbrack ^2
-{{m_{\sigma}^2}\over{32\kappa\yo1}}
{\lbrack tr(\Sigma\Sigma ^{\dagger}) \rbrack ^2}
\log\lbrack{{tr(\Sigma\Sigma ^{\dagger}) }/{2\yo1}}\rbrack +...\cr}
\EQN rcqm $$
\autoparens
As the conformal anomaly is turned off
the interactions with singularities in
field space in the symmetric phase should vanish, requiring $\kappa
,\ga\rightarrow 1$. On the broken symmetry side of the phase
transition, the large-$\nc$ algebraic sum rules, \Eq{aw} and \Eq{sc},
are satisfied with precisely these values in what we call the
``dilaton limit'', defined by ${m_{\sigma}}\rightarrow 0$, and $\kappa
=\ga =1$.
In the dilaton limit (dropping the primes on the $\vec\pi$ and $\sigma$
fields) the lagrangian simplifies to

$$
\eqalign{
&{\cal L}=i{\bar Q}\slashchar D Q
            -{1\over{2}}tr(\roke )
      +{1\over 2}{\pmu}\vec\pi{\pmuu}\vec\pi +
     {1\over 2}{\pmu}\sigma{\pmuu}\sigma
       -{m\over{f_{\pi}}}{\bar Q}\lbrack\sigma -i{\gamma
             _5}{\vec{\pi}\cdot\vec\tau}\rbrack Q\cr
&+{{m_{\sigma}^2}\over{16\yo1}}(\ssd )^2
-{{m_{\sigma}^2}\over{8\yo1}}
{(\ssd )^2}\log\lbrack{{(\ssd )}/{\yo1}}\rbrack+...\cr}
\EQN rcqm2 $$
This is our main result.
The effective theory of constituent quarks
is renormalizable with chiral symmetry breaking induced by
a generalized Coleman-Weinberg potential\ref{cw}.  For our purposes,
renormalizability means that all quanta necessary to ensure
good asymptotic behaviour of scattering amplitudes are already present
in a well defined lagrangian$^5$\vfootnote5{
By well defined we mean that the masses of all quanta in the
effective theory are substantially less than the chiral symmetry breaking
scale.}.
It is amusing that the above lagrangian follows directly
{}from Weinberg's ``theorem'' as the only {\it known} field theory that
is ``equivalent'' (in the manner described above) to QCD.
The emergence of a low-lying scalar in the large-$\nc$ limit is
not at odds with the quasiparticle philosophy inherent to the chiral
quark picture. Moreover, since the scalar is a Goldstone boson,
one can systematically 	include corrections to the dilaton limit.
The scenario advocated here is much in the spirit of the
``vector limit'' of \Ref{geo}.

Perhaps the most convincing support for the point of view advocated in
this letter lies in the study of chiral symmetry restoration at finite
temperature.  In the dilaton limit, the zero temperature effective
theory of constituent quarks is very near the chiral symmetry breaking
phase transition. As temperature approaches its critical value,
the $\x_{i}$ become true
symmetry generators$^6$\vfootnote6{Algebraic realizations and the
chiral symmetry breaking phase transition are discussed in
\Ref{us}.},
and the important modes for
discussing chiral symmetry restoration are the quarks and the
``Goldstone quartet''.
In fact, it has been shown\ref{hk}\ref{gock} that simple constituent
quark models with elementary
$\pi$ and $\sigma$ modes reproduce the pattern of chiral symmetry
restoration observed in finite temperature lattice Monte Carlo calculations.
(Extension of our model to finite temperature is in progress.)

What are the phenomenological implications of our effective
lagrangian? Presumably away from the large-$\nc$ limit the dilaton
mass increases; of course a substantial dilaton mass is already
achieved by inclusion of explicit chiral symmetry breaking effects.
As mentioned earlier, the philosophy of the chiral quark model is
that, except for long-range effects that can be simulated by a
confining potential or a bag, the quark-gluon coupling is small and
allows for a perturbative treatment (the first term of which splits
the delta from the nucleon).  Because the lagrangian \Eq{rcqm2}
results from a derivative expansion, leading order S-matrix elements
consist of the sum of tree graphs involving quarks and the Goldstone
bosons only.  If we further assume that confinement effects can be
neglected for most nucleon structure purposes, the system of coupled
differential equations that follows is known to possess
non-topological soliton solutions.  It is found\ref{bb} that these
baryonic solutions are essentially determined by the coupling of the
quark to the Goldstone quartet, being insensitive to details of the
effective potential; in particular, a rather low sigma mass can be
tolerated while still producing a realistic soliton mass.
Generically, the main drawback of such models is seen from the crudest
non-relativistic approximation: $\ga$ has to be smaller than one in
order to obtain the correct nucleon axial coupling. Of course $\ga$ is
shifted away from unity by loop effects; it is encouraging that for
large values of the dilaton mass $\ga$ indeed decreases\ref{sp1}.  The
``dilated'' chiral quark model also provides at least a partial
explanation of the successes of constituent quark models of nuclear
forces\ref{fa}, where the exchange of a low-lying scalar excitation is
the most successful way of generating the necessary intermediate-range
attraction.

It has long been known that there is a cryptic connection between
renormalizability and conformal invariance\ref{ccj}. We find it
particularly interesting that conformal invariance and chirality
conspire to yield good asymptotic behaviour in an effective theory in
which both symmetries are spontaneously broken$^7$\vfootnote7{
This would appear to be related to the need for a particle with the
properties of the Higgs boson in gauge field theories with
spontaneous symmetry breakdown.}. In the
context of chiral quarks, we have made use of large-$\nc$ algebraic
sum rules to argue that the absence of confining effects implies the
relevance of conformal invariance, with the elementary dilaton field
completing the chiral multiplet and rendering the effective theory
renormalizable.  It is tempting to speculate that there exists a deep
connection between conformal invariance and chiral symmetry breaking.

\vskip0.25in
We are grateful to the ECT* and to the organizers of the Workshop on
Chiral Symmetry in Hadrons and Nuclei, M.~Rho and W.~Weise, for
hospitality while part of this work was completed.  Discussions with
C.B.~Chiu, P.B.~Pal, S.~Varma, and S.~Weinberg are gratefully
acknowledged by one of us (S.R.B).  This work was supported in part by
the Department of Energy Grants DE-FG05-85ER40200 and
DE-FG06-88ER40427.
\nosechead{References}                         
\ListReferences
\end